\documentclass[aps,pre,twocolumn]{revtex4}

\usepackage{amsmath}
\usepackage{amssymb}
\usepackage{graphicx}
\usepackage{dcolumn}
\usepackage{bm}

\newcommand{\eps}{\varepsilon}

\newcommand{\prt}{\partial}

\begin{document}

\title{
Landau-Khalatnikov problem in relativistic hydrodynamics}

\author{ A. M. Kamchatnov}
\affiliation{Institute of Spectroscopy,
Russian Academy of Sciences, Troitsk, Moscow, 108840, Russia}

\date{\today}

\begin{abstract}
An alternative approach to solving the Landau-Khalatnikov problem on
one-dimensional stage of expansion of hot hadronic matter created in collisions of
high-energy particles or nuclei is suggested. Solving the relativistic
hydrodynamics equations by the Riemann method yields a representation for
Khalatnikov's potential which satisfies explicitly the condition of symmetry
of the matter flow with respect to reflection in the central plane of the
initial distribution of matter. New exact relationships are obtained for
evolution of the density of energy in the center of the distribution and for
laws of motion of boundaries between the general solution and the rarefaction waves.
The rapidity distributions are derived in the Landau approximation with account of the
pre-exponential factor.
\end{abstract}

\maketitle

\section{Introduction}

In 1950 E.~Fermi supposed \cite{fermi-50,fermi-53} that in collisions of high-energy
particles so many new elementary particles are created that their mean free path
becomes much smaller than the size of the whole cloud of created nuclear (hadronic) matter.
As a result, the thermal equilibrium is reached very fast and, hence, energy
distribution of particles flying out of the cloud can be found with the use of
statistical mechanics formulas. Soon after that, I.~Ya.~Pomeranchuk \cite{pomeranchuk}
remarked that such a matter must first expand with relativistic velocities and the
real outgoing particles are formed when the temperature drops down the value about their mass.
L.~D.~Landau showed in Ref.~\cite{landau-53} that in collisions of ultra-relativistic
particles the arising cloud must have a form of a thin disk due to Lorentz contraction
and, consequently, the initial stage of hydrodynamic flow is mainly one-dimensional.
If we denote by $2l$ the initial thickness of the disk and take the $x$-axis in the
direction normal to the disk, then equations of relativistic hydrodynamics are
simplified in the limit of ultra-relativistic flow velocities $v$ ($c-v\ll c$, where
$c$ is the speed of light. Landau found the asymptotic solution of these equations for
$\ln(t/l)\gg1$, $\ln[(ct-x)/l]\gg1$ with logarithmic accuracy what permitted him to
make estimates of typical parameters of the flow. In the next years, the Landau theory
formed a basis for theoretical description of multiple particles production in high-energy
collisions of elementary particles and atomic nuclei and this approach was confirmed
in experiments (see, e.g., \cite{wong,flork}).

Exact solution of relativistic hydrodynamics equations for the one-dimensional flow
was obtained by I.~M.~Khalatnikov in Ref.~\cite{khal-54}. He formulated the problem
in the following way. Let at the initial moment $t=0$ the matter be contained in the
slab $-l\leq x\leq l$ and have the temperature $T_0\gg mc^2$, where $m$ is a typical
mass of the matter constituents (say, of pions). For such ultra-relativistic
temperatures it is natural to assume that the equation of state
\begin{equation}\label{2-1}
  p=\eps/3,
\end{equation}
takes place, where $p$ is the pressure and $\eps$ is the energy density. In the case of such
an equation of state the sound velocity is equal to (see \cite{LL-6})
\begin{equation}\label{2-3}
  c_s=c\sqrt{\frac{\prt p}{\prt\eps}}=\frac{c}{\sqrt{3}},
\end{equation}
and it does not depend on the parameters of the medium. At the very initial stage of evolution,
two self-similar rarefaction waves, centered around the points $x=\pm l$, start to propagate
from the edges of the initial distribution. At the moment of time $t_c=l/c_s=\sqrt{3}\,l/c$
they collide in the center of the matter distribution at $x=0$ and after this moment the
region of the general solution arises between the rarefaction waves. Thus, the general
solution must satisfy the boundary conditions which correspond to matching with the
rarefaction waves. Khalatnikov proved in Ref.~\cite{khal-54} an important theorem that
one-dimensional relativistic flow is always potential and with the use of the hodograph
transform he reduced the equation for the potential to a linear second order differential
equation with constant coefficients. This equation is often called in the context of
relativistic hydrodynamics as {\it Khalatnikov equation}. It is equivalent mathematically
to the so-called ``telegraph equation'' which can be solved in the case of the initial
value problems with the use of the Laplace transform \cite{pb-52}. By means of skillful
calculations, Khalatnikov obtained the exact solution which satisfies the necessary
boundary conditions and showed that it reproduces the Landau solution in the appropriate
approximation (see also \cite{bl-55}). More detailed study of the Khalatnikov solution was
later performed in other papers \cite{milekhin,cll-74,wong-14}.

In spite of beauty of the Khalatnikov solution, it is represented in the form which
does not demonstrate explicitly the symmetry of the flow with respect to reflection in
the plane $x=0$ what hinders in some respects its investigation. Besides that, the method
used in Ref.~\cite{khal-54} is applicable, apparently, to the systems with constant
sound velocities and it hardly can be generalized on situations when the equation for
the potential in the hodograph plane has variable coefficients. In this paper, we use
for solving the Landau-Khalatnikov problem the Riemann method \cite{riemann,sommer-50,kgs-70}
which can be applied to a wider class of equations of state. In particular, the problem
of expansion of Bose-Einstein condensate released from a ``box-like'' trap has been
solved recently in Ref.~\cite{ik-19}. The found here form of the solution reflects
explicitly the symmetry of the flow and with its use we have found the rapidity
distribution with account of the pre-exponential factor as well as the exact expressions
for the time dependence of the energy density in the center of the space distribution and
the paths of the boundaries between the general solution and the rarefaction waves.
Comparison of two forms of the solution demonstrates their physical equivalence.

\section{Relativistic hydrodynamics}

From now on, to simplify formulas, we accept the system of units in which the speed of
light $c$ and the initial temperature $T_0$ in the slab are equal to unity. At first
we shall formulate the main fact from relativistic hydrodynamics to introduce the
notation and necessary definitions.

\subsection{Main equations}

As is known \cite{LL-6,anile-89}, equations of relativistic hydrodynamics are contained in
conservation laws of a relativistic flow which in case of a one-dimensional flow have
the form
\begin{equation}\label{4-4}
  \frac{\prt T^{00}}{\prt t}+\frac{\prt T^{01}}{\prt x}=0,\qquad
\frac{\prt T^{01}}{\prt t}+\frac{\prt T^{11}}{\prt x}=0,
\end{equation}
where the components of the energy-momentum tensor $T^{\mu\nu}$ are equal to
\begin{equation}\label{4-5}
\begin{split}
  &T^{00}=(\eps+p)u^0u^0-p,\quad T^{01}=(\eps+p)u^0u^1,\\
  &T^{11}=(\eps+p)u^1u^1+p,
  \end{split}
\end{equation}
and $u^{\mu}$ is the vector of 4-velocity
\begin{equation}\label{4-6}
  (u^0,u^1)=\left(\frac1{\sqrt{1-v^2}},\frac{v}{\sqrt{1-v^2}}\right)=(\cosh y,\sinh y),
\end{equation}
where we have introduced the rapidity $y$ related with the flow velocity $v$ by the
equation $v=\tanh y$, so that Lorentz transformation corresponds to a hyperbolic
rotation by the ``angle'' $y$.  After substitution of Eq.~(\ref{4-6}) into
Eq,~(\ref{4-5}) we arrive with account of definition of the sound velocity
$c_s^2=\prt p/\prt\eps$ to the system of equations
\begin{equation*}
  \begin{split}
&(\cosh^2y+c_s^2\sinh^2y)\frac{\prt\eps}{\prt t}+2(\eps+p)\sinh y\cosh y\,\frac{\prt y}{\prt t}\\
&+(1+c_s^2)\sinh y\cosh y\,\frac{\prt\eps}{\prt x}\\
&+(\eps+p)(\sinh^2y+\cosh^2y)\frac{\prt y}{\prt x}=0,\\
&(1+c_s^2)\sinh y\cosh y\,\frac{\prt\eps}{\prt t}+(\eps+p)(\sinh^2 y+\cosh^2y)\frac{\prt y}{\prt t}\\
&+(\sinh^2y+c_s^2\cosh^2y)\frac{\prt\eps}{\prt x}\\
&+2(\eps+p)\sinh y\cosh y\,\frac{\prt y}{\prt x}=0.
\end{split}
\end{equation*}
Characteristic velocities of these first order differential equations
Х\begin{equation}\label{4-8}
  v_{\pm}=\frac{v\pm c_s}{1\pm vc_s}
\end{equation}
have clear physical meaning: speed of signal's propagation is equal to the sum of
flow velocity and the sound velocity according to the relativistic velocity addition
formula and the sound can propagate downstream and upstream. One can easily find the
Riemann invariants corresponding to these characteristic velocities (see, e.g.,
\cite{LL-6,anile-89,kamch-2000}),
\begin{equation}\label{5-9}
  r_{\pm}=y\pm\int_0^{\eps}\frac{c_sd\eps}{\eps+p},
\end{equation}
which in the case of ultra-relativistic equation of state (\ref{2-1}) reduce to
\begin{equation}\label{5-10}
  r_{\pm}=y\pm\sqrt{3}\,\theta,
\end{equation}
where we have introduced instead of the temperature the variable $\theta=\ln T$ and took into
account that Eq.~(\ref{2-1}) leads to the expression
\begin{equation}\label{5-11}
  \eps=T^4
\end{equation}
for the energy density. The flow velocity and the temperature as functions of the Riemann
invariants are given by the formulas
\begin{equation}\label{5-11b}
\begin{split}
  & v=\tanh y=\tanh\left(\frac{r_++r_-}2\right),\\
  & T=e^{\theta}=\exp\left(\frac{r_+-r_-}{2\sqrt{3}}\right).
  \end{split}
\end{equation}
Hydrodynamic equations expressed in terms of the Riemann invariants take simple diagonal form
\begin{equation}\label{5-12}
  \frac{\prt r_{\pm}}{\prt t}+v_{\pm}(r_+,r_-)\frac{\prt r_{\pm}}{\prt x}=0,
\end{equation}
where the characteristic velocities
\begin{equation}\label{5-13}
  v_{\pm}=\frac{\tanh[(r_++r_-)/2]\pm 1/\sqrt{3}}{1\pm\tanh[(r_++r_-)/2]/\sqrt{3}}
\end{equation}
are also expressed in terms of the Riemann invariants.

\subsection{Rarefaction waves}

Equations (\ref{5-12}), (\ref{5-13}) yield at once solutions for rarefaction waves.
Let us consider, for example, the right edge $x=l$ of the initial distribution of
the matter. Since at the initial moment of time the matter is at rest and its temperature
is equal to unity, then at the initial state both Riemann invariants are equal to zero.
Up to the moment $t_c=\sqrt{3}\,l$ of their collision, the rarefaction waves evolve
independently of each other. Therefore, for example, the right rarefaction wave can
depend on the parameter $l$ through the combination $x-l$ only and, consequently,
the Riemann invariants in the right rarefaction wave can depend on the self-similar
variable $(x-l)/t$ only. Then it follows from Eqs.~(\ref{5-12}) that one of the Riemann
invariants must be constant and the variable $(x-l)/t$ must be equal to the
characteristic velocity corresponding to the other Riemann invariant. In the right
rarefaction wave the flow velocity is positive, that is $y>0$, and the temperature
decreases during expansion of the matter, that is $\theta<0$. Hence, in the right
rarefaction wave the Riemann invariant $r_+$ must be constant and equal to its initial value,
\begin{equation}\label{7-17}
  r_+=y+\sqrt{3}\,\theta=0.
\end{equation}
Then from
\begin{equation}\label{7-18}
  \frac{x-l}t=v_-=\frac{v-1/\sqrt{3}}{1-v/\sqrt{3}}
\end{equation}
we obtain the distribution of the flow velocity
\begin{equation}\label{7-19}
  v=\frac{\sqrt{3}\,(x-l)/t+1}{\sqrt{3}+(x-l)/t},
\end{equation}
and from (\ref{5-11}), (\ref{5-11b}) and (\ref{7-17}) it is easy to find the
distribution of the energy density
\begin{equation}\label{7-20}
  \eps=\left(\frac{1-1/\sqrt{3}}{1+1/\sqrt{3}}\cdot
\frac{1-(x-l)/t}{1+(x-l)/t}\right)^{\frac2{\sqrt{3}}}.
\end{equation}
At the boundary with vacuum, where $\eps\to0$, the matter moves to the right with
the speed of light, what is natural since the matter consists of particles with the
thermal velocities equal in our approximation to the speed of light. Into the slab
the rarefaction wave propagates with the sound velocity $(x-l)/t=-1/\sqrt{3}$ and
at the boundary with the quiescent matter we have $\eps=1$. In a similar way one can
build the solution for the left rarefaction wave which depends on the self-similar
variable $(x+l)/t$. The Riemann invariant $r_-=y-\sqrt{3}\,\theta=0$ is constant within
it what corresponds to the negative flow velocity $y<0$.

\subsection{Khalatnikov equation}

After collision of the rarefaction waves at the moment $t_c=\sqrt{3}\,l$ in the center $x=0$
the region of the general solution appears between them, where both Riemann invariants
$r_{\pm}$ change with $x$ and $t$, and finding the corresponding solution of 
the hydrodynamic equations is a more
complicated problem than finding the rarefaction waves solutions.

Khalatnikov showed in Ref.~\cite{khal-54} that one can obtain from Eqs.~(\ref{4-4}) that
\begin{equation}\label{5-13b}
  \frac{\prt (Tu^1)}{\prt t}+\frac{\prt (Tu^0)}{\prt x}=0,
\end{equation}
which means that we can introduce such a potential $\phi=\phi(x,t)$ that
$$
d\phi=T\cosh y\cdot dt-T\sinh y\cdot dx.
$$
Introducing also the light cone variables $x_{\pm}=t\pm x$ and considering the
Riemann invariants $r_{\pm}$ as independent variables, we can rewrite this
differential in the form
\begin{equation}\nonumber
  d\phi=A(r_+,r_-)dx_-+B(r_+,r_-)dx_+,
\end{equation}
where
\begin{equation}\label{6-14}
\begin{split}
&  A(r_+,r_-)=\frac12e^{\theta+y}\\
&=\frac12\exp\left[\frac12\left(1+\frac1{\sqrt{3}}\right)r_++
\frac12\left(1-\frac1{\sqrt{3}}\right)r_-\right],\\
& B(r_+,r_-)=\frac12e^{\theta-y}\\
&=\frac12\exp\left[-\frac12\left(1-\frac1{\sqrt{3}}\right)r_+-
\frac12\left(1+\frac1{\sqrt{3}}\right)r_-\right].
\end{split}
\end{equation}
Following Khalatnikov, we make Legendre transformation
\begin{equation*}
  W=\phi-Ax_--Bx_+,
\end{equation*}
where the variables $W,x_-,x_+$ are considered as functions of the Riemann
invariants, which corresponds to the hodograph transform known in compressible gas dynamics
(see, e.g., \cite{LL-6,kamch-2000}). As a result, we obtain
\begin{equation*}
\begin{split}
dW&=\frac12\left[-\left(1+\frac1{\sqrt{3}}\right)Ax_-+\left(1-\frac1{\sqrt{3}}\right)Bx_+\right]dr_+\\
&+\frac12\left[-\left(1-\frac1{\sqrt{3}}\right)Ax_-+\left(1+\frac1{\sqrt{3}}\right)Bx_+\right]dr_-,
\end{split}
\end{equation*}
so that
\begin{equation*}
\begin{split}
&  \frac12\left[-\left(1+\frac1{\sqrt{3}}\right)Ax_-+
\left(1-\frac1{\sqrt{3}}\right)Bx_+\right]=\frac{\prt W}{\prt r_+},\\
&  \frac12\left[-\left(1-\frac1{\sqrt{3}}\right)Ax_-+
\left(1+\frac1{\sqrt{3}}\right)Bx_+\right]=\frac{\prt W}{\prt r_-}.
\end{split}
\end{equation*}
Thus, if the function $W(r_+,r_-)$ is found, then $x,t$ are expressed in terms
of the Riemann invariants by the formulas
\begin{equation}\label{6-16}
  \begin{split}
  \begin{split}
 x=\sqrt{3}\,e^{-\theta}&\Big\{\left(\frac{\cosh y}{\sqrt{3}}+\sinh y\right)\frac{\prt W}{\prt r_-}\\
&+\left(\frac{\cosh y}{\sqrt{3}}-\sinh y\right)\frac{\prt W}{\prt r_+}\Big\},\\
 t=\sqrt{3}\,e^{-\theta}&\Big\{\left(\frac{\sinh y}{\sqrt{3}}+\cosh y\right)\frac{\prt W}{\prt r_-}\\
&+\left(\frac{\sinh y}{\sqrt{3}}-\cosh y\right)\frac{\prt W}{\prt r_+}\Big\},
\end{split}
\end{split}
\end{equation}
which yield the solution in implicit form.

So far we have solved formally the equation (\ref{5-13b}) by introduction of the
potential $W=W(r_+,r_-)$. To get the equation for $W$, we can use as a second
equation of relativistic hydrodynamics the entropy conservation law. If the
entropy density is denoted by $\sigma$, then this law reads (see \cite{LL-6}):
\begin{equation}\label{8-21}
  \frac{\prt(\sigma u^0)}{\prt t}+\frac{\prt(\sigma u^1)}{\prt x}=0.
\end{equation}
In case of ultra-relativistic equation of state we have
\begin{equation}\label{8-22}
  \sigma=\frac43T^3=\frac43\exp\left[\frac{\sqrt{3}}2(r_+-r_-)\right].
\end{equation}
We multiply Eq.~(\ref{8-21}) by the Jacobian $\prt(x,t)/\prt(r_+,r_-)$ and
with account of Eqs.~(\ref{8-22}) and (\ref{5-11b}) we obtain the equation
\begin{equation}\label{8-23}
  \frac{\prt^2W}{\prt r_+\prt r_-}-\frac1{2\sqrt{3}}
\left(\frac{\prt W}{\prt r_+}-\frac{\prt W}{\prt r_-}\right)=0.
\end{equation}
Written, instead of $r_{\pm}$, in terms of the variables $y,T$, it was
obtained by Khalatnikov~\cite{khal-54}. It is mathematically equivalent to
the so-called ``telegraph equation'' which can be solved by well developed
mathematical methods (see, e.g., \cite{pb-52,sommer-50,kgs-70}). Some classes
of such solutions are used in the theory of multiple particles production
(see, e.g., Refs.~\cite{bps-08,ps-11}). In the case of the Landau-Khalatnikov
problem on the slab expansion with specific boundary conditions, the Riemann
method seems to be quite convenient, and we shall use it in the next section.

\section{Landau-Khalatnikov problem}

\subsection{Boundary conditions}

\begin{figure}[t]
\begin{center}
\includegraphics[width=8cm]{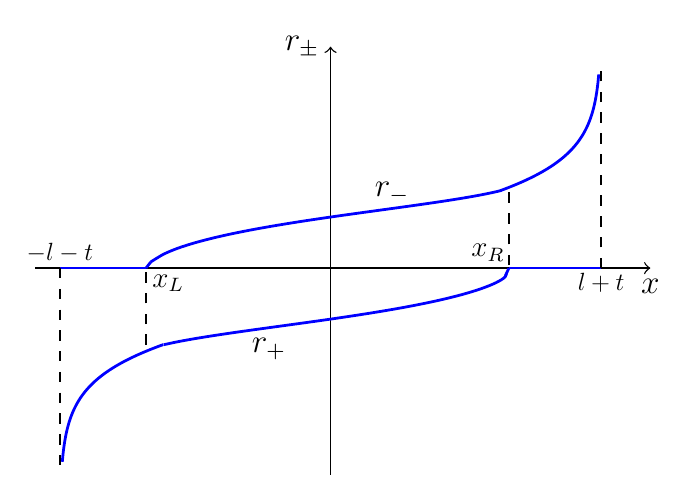}
\caption{ Plots of Riemann invariants $r_{\pm}$ as functions of $x$ at fixed
moment of time $t>\sqrt{3}\,l$; $x_L$ and $x_R$ denote the boundaries of the general solution
and $-l-t$ и $l+t$ denote the boundaries of the left and right rarefaction waves
with vacuum.
}
\label{fig1}
\end{center}
\end{figure}

The potential $W$, which defines the solution by the formulas (\ref{6-16}), must
satisfy Eq.~(\ref{8-23}) and the boundary conditions which follow from matching
with the rarefaction waves at the corresponding edges. If Eqs.~(\ref{6-16}) are
solved with respect to the derivatives $\prt W/\prt r_{\pm}$, then the solution can
be written in the form
\begin{equation}\label{9-24}
  \begin{split}
& \frac{\prt W}{\prt r_-}=\frac{e^{\theta}}2\left(\cosh y-\frac{\sinh y}{\sqrt{3}}\right)
\left(x-\frac{\tanh y-1/\sqrt{3}}{1-\tanh y/\sqrt{3}}\,t\right),\\
& \frac{\prt W}{\prt r_+}=\frac{e^{\theta}}2\left(\cosh y+\frac{\sinh y}{\sqrt{3}}\right)
\left(x-\frac{\tanh y+1/\sqrt{3}}{1+\tanh y/\sqrt{3}}\,t\right).
\end{split}
\end{equation}
Plots of the Riemann invariants as functions of the coordinate $x$ at fixed moment of time
$t$ are shown in Fig.~\ref{fig1}. The Riemann invariants $r_+$ and $r_-$ vanish
at the right and left edges of the general solution, correspondingly.
At these edges $x$ and $t$ satisfy the rarefaction waves solutions (see Eq.~(\ref{7-18})),
that is
\begin{equation}\label{9-25}
\begin{split}
 & x-\frac{\tanh y-1/\sqrt{3}}{1-\tanh y/\sqrt{3}}\,t=l \quad\text{at}\quad r_+=0,\\
  & x-\frac{\tanh y+1/\sqrt{3}}{1+\tanh y/\sqrt{3}}\,t=-l \quad\text{at}\quad r_-=0.
\end{split}
\end{equation}
Substitution of these formulas into Eqs.~(\ref{9-24}) yields the boundary conditions
in the form
\begin{equation}\label{9-26}
  \begin{split}
 \left.\frac{\prt W}{\prt r_-}\right|_{r_+=0}=\frac{l}4&
 \Big\{\left(1-\frac1{\sqrt{3}}\right)e^{\frac12\left(1-\frac1{\sqrt{3}}\right)r_-}+\\
&+\left(1+\frac1{\sqrt{3}}\right)e^{-\frac12\left(1+\frac1{\sqrt{3}}\right)r_-}\Big\},\\
 \left.\frac{\prt W}{\prt r_+}\right|_{r_-=0}=-\frac{l}4&
 \Big\{\left(1+\frac1{\sqrt{3}}\right)e^{\frac12\left(1+\frac1{\sqrt{3}}\right)r_+}+\\
&+\left(1-\frac1{\sqrt{3}}\right)e^{-\frac12\left(1-\frac1{\sqrt{3}}\right)r_+}\Big\},
\end{split}
\end{equation}
Since the potential $W$ is defined up to an additive constant, we can fix its value
$W(0,0)=0$ in the origine of the hodograph plane $(r_+,r_-)$. Then integration of
Eqs.~(\ref{9-26}) gives the final form of the boundary conditions along the boundaries
between the general solution and the rarefaction waves:
\begin{equation}\label{9-27}
\begin{split}
  & W(0,r_-)=\frac{l}2e^{\frac{r_-}{2\sqrt{3}}}\sinh\frac{r_-}2,\\
& W(r_+,0)=-\frac{l}2e^{-\frac{r_+}{2\sqrt{3}}}\sinh\frac{r_+}2.
\end{split}
\end{equation}

\begin{figure}[t]
\begin{center}
\includegraphics[width=8cm]{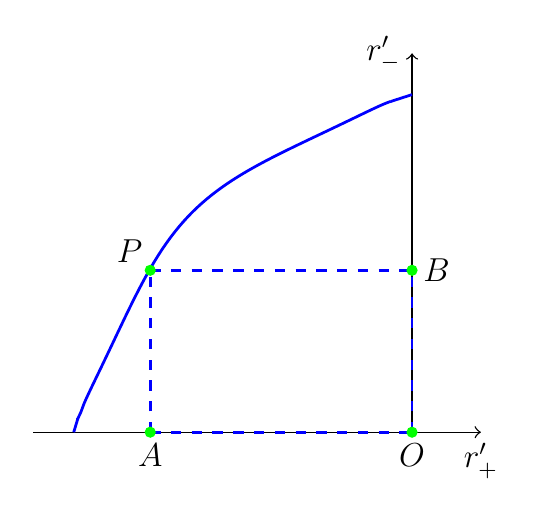}
\caption{Plot of Riemann invariants $r_{\pm}$ in the hodograph plane
$(r_+^{\prime},r_-^{\prime})$ for the general solution at a fixed moment of time.
}
\label{fig2}
\end{center}
\end{figure}

\subsection{Riemann's method}

So, we have to solve Eq.~(\ref{8-23}) with the boundary conditions (\ref{9-27})
prescribed on the characteristics $r_+=0$ and $r_-=0$ of this equation.
In his fundamental paper \cite{riemann}, B.~Riemann gave the following method
of solving this problem. If we are interested in the value of the function
$W$ at the point $P=(r_+,r_-)$ (see Fig.~\ref{fig2}), then we should draw
in the hodograph plane $(r_+',r_-')$ from the point $P$ two characteristics
$PA$ $(r_+'=r_+=\mathrm{const})$ and $PB$ $(r_-'=r_-=\mathrm{const})$, which
together with the characteristics $AO$ and $OB$ with known values of $W(r_+',0)$
and $W(0,r_-')$ along them form a closed contour $\mathcal{C}=PAOB$ in this plane.
Since the symbols $(r_+,r_-)$ denote now the coordinates of the ``observation point''
in the hodograph plane, we have introduced the notation $(r_+',r_-')$ for varying
along the contour $\mathcal{C}$ coordinates. We define in this plane such a vector
$(V,U)$, that the integral $\int_{\mathcal{C}}(Vdr_+'+Udr_-')=0$ vanishes. The
components $(V,U)$ depend here on the function $W$ satisfying Eq.~(\ref{8-23})
and also on another function $R=R(r_+',r_-';r_+,r_-)$ on which we can impose
some appropriate conditions without changing the zero value of the above integral.
Riemann showed that if one demands, first, that the function $R$ satisfies the conjugated
equation which in our case reads
\begin{equation}\label{11-28}
  \frac{\prt^2R}{\prt r_+'\prt r_-'}+\frac1{2\sqrt{3}}
\left(\frac{\prt R}{\prt r_+'}-\frac{\prt R}{\prt r_-'}\right)=0,
\end{equation}
second, it satisfies the boundary conditions
\begin{equation}\label{11-29}
  \begin{split}
& \frac{\prt R}{\prt r_+'}-\frac{R}{2\sqrt{3}}=0\quad\text{along}\quad PB;\\
& \frac{\prt R}{\prt r_-'}+\frac{R}{2\sqrt{3}}=0\quad\text{along}\quad PA,
\end{split}
\end{equation}
and, third, we fix its value at the point $P$,
\begin{equation}\label{11-30}
  W(r_+,r_-;r_+,r_-)=1,
\end{equation}
then the value of $W$ at the point $P=(r_+,r_-)$ will be given by the formula
\begin{equation}\label{11-31}
  W(P)=\frac12(RW)_A+\frac12(RW)_B+\int_A^OVdr_+' +\int_O^BUdr_-',
\end{equation}
where $A=(r_+,0)$, $B=(0,r_-)$, $O=(0,0)$ and $U$, $V$ are defined in terms of
the boundary values (\ref{9-27}) by the expressions
\begin{equation}\label{12-32}
  \begin{split}
& U=\frac12\left({W}\frac{\prt R}{\prt r_+'}-R\frac{\prt {W}}{\prt r_+'}\right)
-\frac{{W}R}{2\sqrt{3}},\\
& V=\frac12\left(R\frac{\prt {W}}{\prt r_-'}-{W}\frac{\prt R}{\prt r_-'}\right)
-\frac{{W}R}{2\sqrt{3}}.
\end{split}
\end{equation}
Although it seems that to solve Eq.~(\ref{11-28}) for $R$ is not easier than to
solve Eq.~(\ref{8-23}) for $W$, the boundary conditions (\ref{11-29}) are now so
simple that the function $R$ can be found in our case without much difficulty.
Since the equation of state (\ref{2-1}) is similar to the non-relativistic
isothermic equation of state $p=c_s^2\rho$ with constant isothermic sound velocity
$c_s$ which does not depend on the density of gas, and the corresponding function
$R$ was found by Riemann himself in Ref.~\cite{riemann}, we can adjust his result
to our case to obtain
\begin{equation}\label{12-33}
\begin{split}
  R(r_+',r_-';r_+,r_-)& = \exp\left[\frac{(r_+'-r_+)+(r_--r_-')}{2\sqrt{3}}\right]\\
& \times I_0\left(\sqrt{(r_+'-r_+)(r_--r_-')/3}\right),
\end{split}
\end{equation}
where $I_0(z)$ is the Bessel function of complex argument (see, e.g., Ref.~\cite{ww}).

When we have the Riemann formula (\ref{11-31}) with known expression (\ref{12-33}) for the
Riemann function $R$, then it is not difficult to get the expression for the potential $W$.
To this end, we notice first, that substitution of (\ref{12-32}) into (\ref{11-31}) and
elementary integration by parts cast the Riemann formula to
\begin{equation}\label{12-34}
\begin{split}
  W(P)=& -\int_A^OR\left( \frac{\prt W}{\prt r_+'}+\frac{W}{2\sqrt{3}}\right)dr_+'+\\
& +\int_O^BR\left(\frac{\prt{W}}{\prt r_-'}-\frac{W}{2\sqrt{3}}\right)dr_-'.
\end{split}
\end{equation}
Then substitution of the boundary expressions (\ref{9-27}) for $W$ yields after simple
transformations
\begin{equation}\label{12-35}
  \begin{split}
W(r_+,r_-)&=\frac{l}2\exp\left(\frac{r_--r_+}{2\sqrt{3}}\right)\\
& \Bigg\{
\int_{r_+}^0\left(\cosh\frac{r}2+\frac2{\sqrt{3}}\sinh\frac{r}2\right)e^{\frac{r}{\sqrt{3}}}\\
&\times 
I_0\left(\sqrt{\frac{(r-r_+)r_-}3}\right)dr\\
&+\int_0^{r_-}\left(\cosh\frac{r}2-\frac2{\sqrt{3}}\sinh\frac{r}2\right)e^{-\frac{r}{\sqrt{3}}}\\
&\times 
I_0\left(\sqrt{\frac{-r_+(r_--r)}3}\right)dr\Bigg\}.
\end{split}
\end{equation}
This formula together with Eqs.~(\ref{6-16}) defines implicitly the dependence of the Riemann
invariants $r_{\pm}$ on $x$ and $t$, and with the help of Eqs.~(\ref{5-11b}) we find the
dependence of the physical variables on coordinate and time. The above expression for $W$
seems somewhat more complex than the corresponding expression for the potential in
Khalatnikov's solution \cite{khal-54} (see below Section~4), but it is invariant with respect to
the transformation $r_+\to-r_-$, $r_-\to-r_+$, which corresponds to the change of the sign
of the flow velocity which means the symmetry of the flow with respect to reflection in the
plane $x=0$. Such a symmetry is absent in Khalatnikov's expression for the potential.
Let us infer some consequences from the obtained solution.

\subsection{Initial stage of evolution of the general solution region}

For small time $t-t_c\ll t_c$ after the collision moment $t_c=l\sqrt{3}$ of the rarefaction
waves, the absolute values of the Riemann invariants $r_{\pm}$ are also small and we can
use the series expansions of the derivatives of $W(r_+,r_-)$,
\begin{equation}\label{13-36}
\begin{split}
&  \frac{\prt W}{\prt r_+}\approx l\left(-\frac12-\frac{r_++r_-}{2\sqrt{3}}
-\frac{r_+^2}8\right),\\
& \frac{\prt W}{\prt r_-}\approx l\left(\frac12-\frac{r_++r_-}{2\sqrt{3}}
+\frac{r_-^2}8\right).
\end{split}
\end{equation}
Then Eqs.~(\ref{6-16}) yield the series expansions
\begin{equation}\label{13-37}
  \begin{split}
 x  \approx & l(r_++r_-)\left(\frac1{2\sqrt{3}}-\frac{5(r_+-r_-)}{24}\right),\\
 t  \approx & l\Big(\sqrt{3}+\frac{r_--r_+}2+\frac{\sqrt{3}}8(r_+^2+r_-^2)-\\
&-\frac{r_+r_-}{2\sqrt{3}}\Big),\qquad |r_+|,r_-\ll1.
\end{split}
\end{equation}
At $t=l\sqrt{3}$ we have $r_+=r_-=0$, as it should be, and during further evolution
we always have $r_+=-r_-$ at $x=0$ (see Fig.~\ref{fig1}).

At the right boundary between the general solution and the rarefaction wave, the
invariant $r_+$ vanishes and substitution of $r_+=0$ yields the law of motion of this
boundary in a parametric form,
\begin{equation}\label{13-38}
\begin{split}
  x_R\approx & lr\left(\frac1{2\sqrt{3}}+\frac{5r}{24}\right),\\
t\approx & l\Big(\sqrt{3}+\frac{r}2
+\frac{\sqrt{3}}8r^2\Big),\qquad r\ll1,
\end{split}
\end{equation}
that is, with the same accuracy,
\begin{equation}\label{13-39}
\begin{split}
   x_R(t)\approx &l\left\{\frac{t}{l\sqrt{3}}-1-4\left(\frac{t}{l\sqrt{3}}-1\right)^2\right\},\\
& \frac{t}{l\sqrt{3}}-1\ll 1.
  \end{split}
\end{equation}
Thus, at the initial stage this boundary propagates with the sound velocity.

In the center of matter distribution with $r_-=-r_+\equiv r$ we obtain
$t\approx l(\sqrt{3}+r+(\sqrt{3}/12)r^2)$, so that the temperature at this point
decreases with time according to the formula
\begin{equation}\label{13-40}
\begin{split}
  T=& e^{-\frac{r}{\sqrt{3}}}\approx 1-\left(\frac{t}{l\sqrt{3}}-1\right)
  +\frac34 \left(\frac{t}{l\sqrt{3}}-1\right)^2,\\
 & \frac{t}{l\sqrt{3}}-1\ll 1.
 \end{split}
\end{equation}

\subsection{Motion of boundaries of the general solution}

\begin{figure}[t]
\begin{center}
\includegraphics[width=8cm]{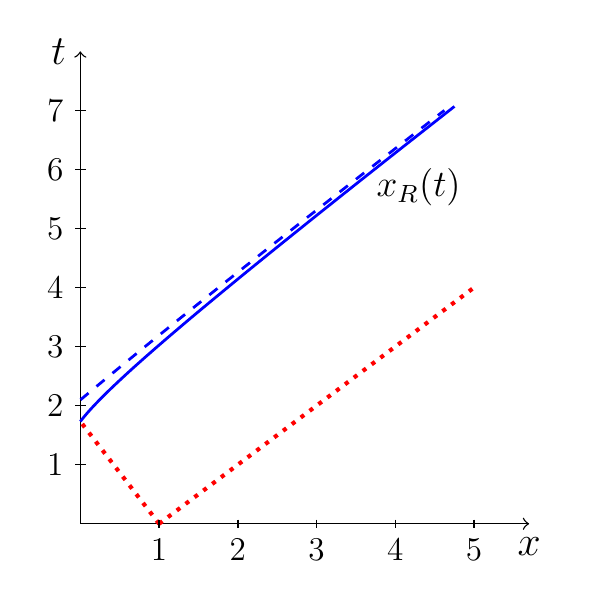}
\caption{The solid line shows the motion of the right boundary $x_R(t)$ according to
the exact formulas (\ref{14-42}) with $l=1$, whereas the dashed line according to the
asymptotic formula (\ref{14-43}). Dotted lines depict the paths of the rarefaction wave
edges: its right edge propagates into vacuum with the light speed and the left edge
into the initial matter distribution with the sound speed $1/\sqrt{3}$ till the
collision moment with the left rarefaction wave in the center at $x=0$.
}
\label{fig3}
\end{center}
\end{figure}

Since the flow is symmetric with respect to the plane $x=0$, it is enough to consider
the motion of the right boundary only $x_R(t)$ where $r_+=0$. At this point the
value of $\left.\prt W/\prt r_-\right|_{r_+=0}$ is already known from the
boundary condition (\ref{9-26}) and a similar limit for $\prt W/\prt r_+$ can
be easily found, so that we get $(r=r_-)$
\begin{equation}\label{14-41}
  \begin{split}
& \left.\frac{\prt W}{\prt r_+}\right|_{r_+=0}=\frac{l}2e^{-\frac{r}{2\sqrt{3}}}
\left(\cosh\frac{r}2+\frac{1}{\sqrt{3}}\sinh \frac{r}2 - 2e^{\frac{r}{\sqrt{3}}}\right),\\
& \left.\frac{\prt W}{\prt r_-}\right|_{r_+=0}=\frac{l}2e^{-\frac{r}{2\sqrt{3}}}
\left(\cosh\frac{r}2-\frac{1}{\sqrt{3}}\sinh \frac{r}2\right).
\end{split}
\end{equation}
Substitution of these formulas into Eqs.~(\ref{6-16}) yields the parametric form for the
law of motion of the right boundary,
\begin{equation}\label{14-42}
\begin{split}
&x_R=l\left\{1+e^{\frac{r}{\sqrt{3}}}\left(\sqrt{3}\,\sinh \frac{r}2-\cosh\frac{r}2\right)\right\},\\
&  t=le^{\frac{r}{\sqrt{3}}}\left(\sqrt{3}\,\cosh\frac{r}2-\sinh \frac{r}2\right).
\end{split}
\end{equation}
For small $r\ll1$ these formulas reproduce, naturally, Eqs.~(\ref{13-38}), and for asymptotically
large time $t/l\gg1$, when $r\gg1$, we obtain
\begin{equation}\label{14-43}
  x\approx l+t-l(\sqrt{3}+1)\left(\frac2{\sqrt{3}-1}\cdot\frac{t}l\right)^{\frac{2-\sqrt{3}}{2+\sqrt{3}}},
\qquad t\gg l.
\end{equation}
As we see, this boundary propagates with velocity close to the light speed and it lags behind the
edge $x=t$ at the distance growing with time. However, the ralative size of the region
occupied by the rarefaction wave decreases with time. It is easy to write down analogous formulas
for motion of the left boundary. The paths of edges of the right rarefaction wave and the
general solution are shown in Fig.~\ref{fig3}.

It is natural to study the time dependence of the energy density at the edge of the general
solution as a function of a proper time $\overline{t}$ on the clock moving with this edge,
\begin{equation}\label{14-44}
\begin{split}
  \overline{t}=\int_{\sqrt{3}\,l}^{t}dt\sqrt{1-v^2}&=\int_0^{r}\frac{dt/dr}{\cosh(r/2)}dr\\
&\approx l\cdot\frac{\sqrt{3}+1}2\exp\left(\frac{r}{\sqrt{3}}\right),
\end{split}
\end{equation}
where $t=t(r)$ is expressed by the second formula (\ref{14-42}). The parameter $r$
equals to $r=-2\sqrt{3}\,\theta$ and, consequently, the energy density $\eps=T^4=e^{4\theta}$
drops down with growth of the proper time as
\begin{equation}\label{15-45}
  \eps\approx\left(1+\frac{\sqrt{3}}2\right)\cdot\left(\frac{l}{\overline{t}}\right)^2.
\end{equation}

\subsection{Evolution of the energy density in the center of the distribution}

In the center of the matter distribution at $x=0$, where $r_-=-r_+\equiv r$, the second
formula (\ref{6-16}) yields after simple transformations
\begin{equation}\label{15-46}
\begin{split}
  t=l\sqrt{3}\,e^{\frac{r}{\sqrt{3}}}\frac{d}{dr}
&\Big\{ e^{\frac{r}{\sqrt{3}}}
\int_0^{r} \left(\cosh\frac{r'}2-\frac{1}{\sqrt{3}}\sinh \frac{r'}2\right)\\
& \times e^{-\frac{r'}{\sqrt{3}}}
I_0\left(\sqrt{\frac{r(r-r')}3}\right)dr'\Big\},
\end{split}
\end{equation}
what determines the dependence of the parameter $r$ on time $t$. Since here
$T=e^{\theta}=e^{-r/\sqrt{3}}$, we find thus the time dependence of the temperature
and of the energy density in the center. At asymptotically large time, when
$t\gg l$ and $r\gg1$, we take into account that the integral in Eq.~(\ref{15-46})
converges at $r'\sim1$ due to the exponential factor $\exp(-r'/\sqrt{3})$ in the
integrand and therefore we can use the asymptotic approximation (see Ref.~\cite{ww})
\begin{equation}\label{15-47}
  I_0(z)\approx\frac{e^z}{\sqrt{2\pi z}},
\end{equation}
for the Bessel function, so that
\begin{equation}\label{15-48}
  I_0\left(\sqrt{\frac{r(r-r')}3}\right)\approx\frac{e^{\frac{r}{\sqrt{3}}}
  \cdot e^{-\frac{r'}{2\sqrt{3}}}}
{\sqrt{2\pi r/\sqrt{3}}},\qquad r\gg r'\sim1,
\end{equation}
and replace the upper limit of integration by infinity. As a result, we obtain
\begin{equation}\label{15-49}
  t\approx l\,\sqrt{\frac{2}{3\pi(-\theta)}}e^{-3\theta},\qquad |\theta|\gg1.
\end{equation}
This equation can be solved with respect to ${-\theta}$ with logarithmic accuracy
and we get
$$
-\theta\approx \frac13\ln\left(\frac{t}l\right),
$$
that is
\begin{equation}\label{15-50}
  T\approx\left(\frac2{\pi\ln(t/l)}\right)^{1/6}\left(\frac{l}t\right)^{1/3},
\end{equation}
which means that the energy density decreases with time as
\begin{equation}\label{15-51}
  \eps=T^4\approx\frac1{(2\pi\ln(t/l))^{2/3}}\cdot\left(\frac{2l}t\right)^{4/3},\qquad t\gg l.
\end{equation}
The law $\eps\sim(l/t)^{4/3}$ was obtained by Landau \cite{landau-53} in the approximation
in which the logarithmically dependent on time factors are neglected.

\subsection{Flow far from the boundaries of the general solution with
rarefaction waves}

\begin{figure}[t]
\begin{center}
\includegraphics[width=8cm]{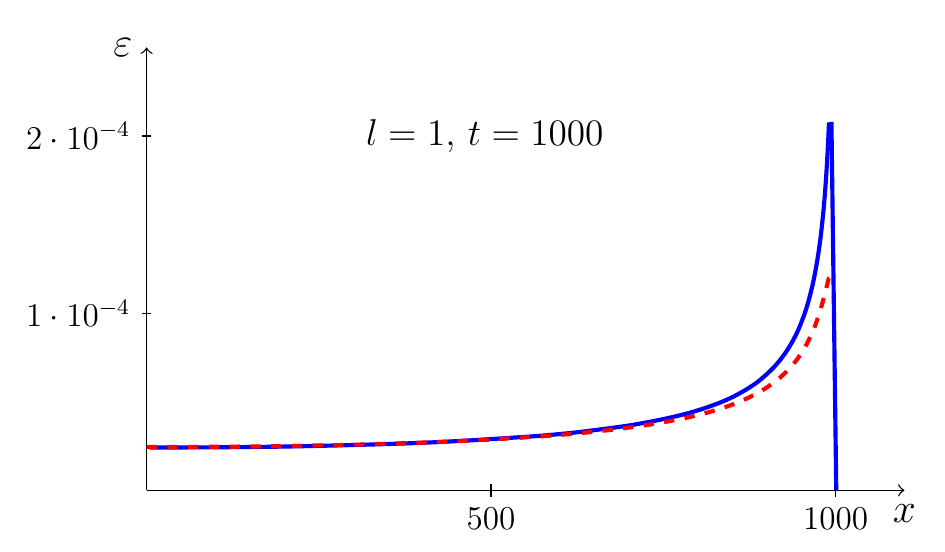}
\caption{Distribution of the energy density $\eps(x)$ on $x>0$ at $t=1000$ for $l=1$
is shown by a solid line for asymptotic solution (\ref{16-52}) complemented by the
rarefaction wave. Dashed line corresponds to the Landau approximation with the
pre-exponential factor in Eq.~(\ref{16-52}) replaced by a constant chosen in such
a way that both distributions coincide at $x=0$.
}
\label{fig4}
\end{center}
\end{figure}

At large time the temperature of matter decreases very much compared with its initial value,
hence $|\theta|=|\ln T|\gg1$. If we are interested in properties of flow far from the
edges of the general solution, where one of the Riemann invariants vanishes, then we
can assume that absolute values of both Riemann invariants $r_{\pm}=y\pm\sqrt{3}\theta$
are large and use the asymptotic formula (\ref{15-47}) for calculation of integrals in
Eq.~(\ref{12-35}) replacing again the limits of integration by  $r_+\to-\infty$,
$r_-\to\infty$. As a result, we arrive at a simple formula for $W$,
\begin{equation}\label{16-52}
\begin{split}
W(r_+,r_-)\approx &l\cdot 3^{3/4}\sqrt{\frac2{\pi}}
\frac{(-r_+r_-)^{1/4}}{r_--r_++4\sqrt{-r_+r_-}}\\
 &\times \exp\left[\frac{(\sqrt{r_-}+\sqrt{-r_+})^2}{2\sqrt{3}}\right].
  \end{split}
\end{equation}
If the pre-exponential factor is omitted, then we return to the solution obtained by
Landau \cite{landau-53,bl-55}. For calulation of derivatives of this expression it is
enough in the main approximation to differentiate the exponent only, since then the
order of the pre-exponential factor does not decrease,
\begin{equation}\label{16-53}
  \begin{split}
\frac{\prt W}{\prt r_+}\approx &-l\frac{3^{1/4}}{\sqrt{2{\pi}}}
\cdot\left(\frac{r_-}{-r_+}\right)^{1/4}\frac{\sqrt{r_-}+\sqrt{-r_+}}{r_--r_++4\sqrt{-r_+r_-}}\\
 &\times \exp\left[\frac{(\sqrt{r_-}+\sqrt{-r_+})^2}{2\sqrt{3}}\right],\\
\frac{\prt W}{\prt r_-}\approx &l\frac{3^{1/4}}{\sqrt{2{\pi}}}
\cdot\left(\frac{-r_+}{r_-}\right)^{1/4}\frac{\sqrt{r_-}+\sqrt{-r_+}}{r_--r_++4\sqrt{-r_+r_-}}\\
 &\times \exp\left[\frac{(\sqrt{r_-}+\sqrt{-r_+})^2}{2\sqrt{3}}\right].
  \end{split}
\end{equation}
Substitution of these formulas into Eqs.~(\ref{6-16}) determines implicitly the dependence
of the Riemann invariants on $x$ and $t$. The Riemann invariants change within the intervals
\begin{equation}\label{16-54}
\begin{split}
  &-r_m\lesssim r_+\lesssim 0,\qquad 0\lesssim r_-\lesssim r_m,\\
  &r_m\approx \frac{2+\sqrt{3}}{2\sqrt{3}}\ln\left(\frac2{\sqrt{3}-1}\cdot\frac{t}l\right),
  \end{split}
\end{equation}
as it can be found from the second formula (\ref{14-42}). If we fix time $t$ and choose some
value of, say, $r_+$ in this interval, then the corresponding value of $r_-$ can be found
from the second equation (\ref{6-16}) and the corresponding value of $x$ is determined by
the first formula (\ref{6-16}). Thus, we can find the values of the Riemann invariants at
fixed $t$ as functions of $x$ which determines the distributions of the physical parameters
according to Eqs.~(\ref{5-11b}). The general character of the flow was determined in
Refs.~\cite{landau-53,bl-55}, where it was shown that the energy is concentrated in the
ultra-relativistic regions of the flow whereas the particles are located mainly in the
region of relatively small velocities. We shall not discuss all that in any details and notice only
that now the distributions can be calculated with account of the pre-exponential factor.
Typical distribution of the energy density is shown in Fig.~\ref{fig4}. Time of evolution
$t=1000$ is chosen large to make the Landau approximation accurate enough with values of
the Riemann invariants about $r_{\pm}\sim5$. However, for not too much values of time the
role of the pre-exponential factor becomes more important which can be noticed in
experimentally measured rapidity distributions of resulting particles.

\subsection{Distributions on rapidities}

Experimentally measured quantity is the distribution of particles on the rapidity
$y$ (see, e.g., \cite{bearden-05}) which is usually compared with the Gauss formula
corresponding to the Landau approximation. Here we derive more accurate formula
with account of the pre-exponential factor.

According to Landau \cite{landau-53}, distributions of particle density coincides with
entropy distributions. Therefore, the rapidity distribution is proportional to the
entropy distribution of each fluid matter particle at the moment of its transition
into real hadrons at the temperature of the order of their mass \cite{milekhin}. In the
proper reference frame of the fluid particle amount of the entropy in the small
layer with thickness $dx'$ is equal to $\sigma dx'=\sigma (u^0)'\cdot dx'$, where
$(u^0)'=1$ is the time-component of the 4-velocity. This expression can be
considered as a product of the entropy flux by the space component of the 4-vector
$dx^i$. In Minkowski geometry, the quantity $(u^0)'dx'$ is an ``area'' element
$A^{\prime 01}=(u^0)'dx'-(u^1)'dt'$, where $(u^{1})'=0$ in the proper reference frame
of the fluid particle. This tensor component is invariant with respect to Lorentz
transformation and, hence, we obtain the expression for the differential of entropy
in an arbitrary reference frame as
\begin{equation}\label{t4-47.37}
  dS=\sigma(u^0dx-u^1dt).
\end{equation}
In the center of inertia reference frame the coordinates $t$ and $x$ are related
with the parameters $\theta$, $y$ by the formulas
\begin{equation}\label{17-55}
  \begin{split}
  t=e^{-\theta}\left(\sinh y\frac{\prt W}{\prt y}-\cosh y\frac{\prt W}{\prt\theta}\right),\\
  x=e^{-\theta}\left(\cosh y\frac{\prt W}{\prt y}-\sinh y\frac{\prt W}{\prt\theta}\right),
  \end{split}
\end{equation}
which are obtained from Eqs.~(\ref{6-16}) by means of the replacements (\ref{5-10}).
Their substitution into Eq.~(\ref{t4-47.37}) with account of
$\sigma\propto T^3=e^{3\theta}$, $u^0=\cosh y$, $u^1=\sinh y$ yields
$$
dS\propto e^{2\theta}\left\{\left(\frac{\prt^2W}{\prt y^2}-\frac{\prt W}{\prt \theta}\right)dy+
\left(\frac{\prt^2W}{\prt\theta\prt y}-\frac{\prt W}{\prt y}\right)d\theta\right\}.
$$
At the moment of formation of real particles one can neglect the change of temperature
and to get for distribution of particles on their rapidities the expression
$$
dN\propto dS\propto e^{2\theta}\left(\frac{\prt^2W}{\prt y^2}-\frac{\prt W}{\prt \theta}\right)dy,
$$
where $dN$ is the number of particles with rapidities within the interval $(y,y+dy)$.
In the asymptotic region far from the boundaries along which $y=\pm\sqrt{3}\,\theta$,
a simple calculation yields
\begin{equation}\label{18-59}
  \begin{split}
  dN=& G(-\theta)\cdot\frac{\theta-\sqrt{\theta^2-y^2/3}}
  {(\theta^2-y^2/3)^{1/4}(\theta-2\sqrt{\theta^2-y^2/3})}\\
  & \times \exp\left[\theta+\sqrt{\theta^2-\frac{y^2}3}\right]dy,
  \end{split}
\end{equation}
where $G(-\theta)$ is a normalization factor defined by the condition
$\int_{\sqrt{3}\theta}^{-\sqrt{3}\theta}dN=1$ and $\theta=\ln T$ corresponds to
the temperature of transformation of hadronic matter to separate particles.
In fact, the dependence of $G(-\theta)$ on $-\theta$ is very weak and it
changes from the value $G(3)\approx0.3597$ to its limiting value
$G(\infty)=(1/2)\sqrt{3/(2\pi)}\approx0.3455$ by 4\% only. For
$-\theta\gg1$, $y\ll|\theta|$ this distribution becomes Gaussian one,
\begin{equation}\label{18-60}
  dN=\frac1{\sqrt{6\pi(-\theta)}}\exp\left(-\frac{y^2}{6(-\theta)}\right)dy,
\end{equation}
in agreement with the experiment at very large collision energy \cite{bearden-05}.
For more moderate energy of collision it may be necessary to use more accurate
formula (\ref{18-59}).

\section{Comparison with the solution in Khalatnikov's form}

I.~M.~Khalatnikov used in Ref.~\cite{khal-54} the reference frame with the right edge
of the initial distribution located at the point $x=0$ (rather than at $x=0$ as we do).
However, the relationships (\ref{17-55}) between the coordinates $x,t$ and parameters
$\theta,y$  were used without translation transformation $x\to x-l$. As a result, the
expression obtained for the potential,
\begin{equation}\label{19-61}
  \widetilde{W}(\theta,y)=l\sqrt{3}\,e^{\theta}\int_{y/\sqrt{3}}^{-\theta}e^{2z}
  I_0\left(\sqrt{z^2-y^2/3}\right)dz,
\end{equation}
is not invariant with repect to change of the sign of the flow velocity $y\to-y$, but
it has simpler form compared with our expression (\ref{12-35}). Written in terms
of the Riemann invariants, the Khalatnikov solution takes the form
\begin{equation}\label{19-62}
\begin{split}
  & \widetilde{W}(r_+,r_-)=  l\sqrt{3}\,e^{\frac{r_+-r_-}{2\sqrt{3}}}\times\\
  &\times \int_{\frac{r_-+r_+}{2\sqrt{3}}}^{\frac{r_--r_+}{2\sqrt{3}}}
  e^{2z} I_0\left(\sqrt{z^2-\frac{(r_+-r_-)^2}{12}}\right)dz.
  \end{split}
\end{equation}
It is evident that the functions $W$ in (\ref{12-35}) and $\widetilde{W}$ in (\ref{19-62})
do not coincide with each other, since they have different symmetry properties. But,
of course, their physical consequences are the same. For example, at the right boundary
between the general solution and the rarefaction wave at $y=-\sqrt{3}\,\theta$ we find 
the derivatives
\begin{equation}\nonumber
  \left.\frac{\prt\widetilde{W}}{\prt y}\right|_{y=-\sqrt{3}\,\theta}=-le^{-\theta},\quad
  \left.\frac{\prt\widetilde{W}}{\prt \theta}\right|_{y=-\sqrt{3}\,\theta}=
  -l\sqrt{3}\,e^{-\theta},
\end{equation}
and their substitution into Eq,~(\ref{17-55}) gives the law of motion of this boundary
coinciding with Eq.~(\ref{14-42}) after taking into account the translational shift
of the origine of the coordinate system to the point $x=l$. At the center of the
matter distribution, where $y=0$, Eq.~(\ref{19-61}) gives the expression 
($r=-\sqrt{3}\,\theta$)
\begin{equation}\label{20-63}
  t=l\sqrt{3}\left\{e^{\frac{2r}{\sqrt{3}}}I_0\left(\frac{r}{\sqrt{3}}\right)-
  \int_0^{\frac{r}{\sqrt{3}}}e^{2z}I_0(z)dz\right\},
\end{equation}
which seems different from Eq.~(\ref{15-46}), but the coincidence of the two
expressions can be easily checked numerically.

It is worth noticing that one cannot calculate the pre-exponential factor by a simple
substitution of the asymptotic formula (\ref{15-47}) into Eq.~(\ref{19-62}) since
now there is no factors in the integrand function which provides convergence of
the integral at finite $z$. Thus, each form of the solution has its own
advantages in studying the flow in this problem.

\section{Conclusion}

In this paper, we have obtained a new form of solution of the Landau-Khalatnikov
problem about expansion of a slab of matter in the relativistic hydrodynamics. 
This problem was formulated first by Landau \cite{landau-53} and he got its
asymptotic solution. The Riemann method \cite{riemann,sommer-50} applied to 
the Khalatnikov equation \cite{khal-54} allows one to get the solution which 
satisfies the necessary boundary condition at the boundaries with the rarefaction
waves and represents explicitly the symmetry of the flow. Although this solution
is physically equivalent to the Khalatnikov solution, its mathematical form has
some advantages and permits one to obtain the asymptotic solution with account of
the pre-exponential factor. We have performed quite detailed study of parameters
of the flow at all stages of its evolution.

\end{document}